\documentclass{elsart_ch}
\usepackage{epsfig,array,amscd}
\usepackage{amsmath,amssymb}
\def\bf{\bfseries}
\include{defines}
\begin{document}

\setcounter{page}{1}

\frenchspacing

\sloppy

\begin{flushright}
CERN-PH-EP/2004-054

October 11, 2004
\end{flushright}
\vspace*{10mm}
\begin{frontmatter}

\title{Measurement of the Radiative $K_{e3}$ Branching Ratio}
\date{}

\collab{NA48 Collaboration}
\author{A.~Lai},
\author{D.~Marras}
\address{Dipartimento di Fisica dell'Universit\`a e Sezione dell'INFN di Cagliari, \\ I-09100 Cagliari, Italy} 
\author{A.~Bevan},
\author{R.S.~Dosanjh},
\author{T.J.~Gershon},
\author{B.~Hay},
\author{G.E.~Kalmus},
\author{C.~Lazzeroni},
\author{D.J.~Munday},
\author{E.~Olaiya\thanksref{threfRAL}},
\author{M.A.~Parker},
\author{T.O.~White},
\author{S.A.~Wotton}
\address{Cavendish Laboratory, University of Cambridge, Cambridge, CB3~0HE, U.K.\thanksref{thref3}}
\thanks[thref3]{Funded by the U.K.\ Particle Physics and Astronomy Research Council}
\thanks[threfRAL]{Present address: Rutherford Appleton Laboratory, Chilton, Didcot, Oxon, OX11~0QX, U.K.}
% ...

\vspace*{\fill}
\end{frontmatter}

\setcounter{footnote}{0}

\newpage

\small\noindent
{\bf Abstract}.

We present 
a measurement of the relative branching ratio of the decay 
$K^0\rightarrow\pi^{\pm}e^{\mp}\nu\gamma$ ($K_{e3\gamma}$) 
with respect to $K^0\rightarrow\pi^{\pm}e^{\mp}\nu$ ($\gamma$) ($K_{e3}+K_{e3\gamma}$) decay.
The result is based on observation of  19 000 $K_{e3\gamma}$
and  $5.6\cdot 10^6$  $K_{e3}$ decays. The value of
the branching ratio is $Br(K^0_{e3\gamma},
E_{\gamma}^*>30~ \rm{MeV},\theta_{e\gamma}^*>20^o)/Br(K^0_{e3})=
(0.964\pm0.008^{+0.011}_{-0.009}) \%$. This result agrees with 
theoretical predictions but
is at variance  with a recently published 
result.


\section{Introduction}

The study of radiative $K_L$ decays can give 
valuable information on the kaon structure.
It allows a good
test of theories describing hadron interactions and decays, 
like Chiral Perturbation Theory (ChPT). 
Here we 
present a study of the radiative 
$K_{e3}$ decay. 


There  are 
two distinct photon components in the radiative $K^0_{e3}$ decays - 
inner bremsstrahlung (IB) and direct emission.
$K^0_{e3}$ decays are mainly sensitive to the IB component because of
the small electron mass $m_e$. 
A big contribution to the rate,
dominated by the IB amplitude, comes from the 
region of small
photon energies $E_{\gamma}^*$
and angles $\theta_{e\gamma}^*$ between the charged lepton
and the photon, with both $E_{\gamma}^*$ and $\theta_{e\gamma}^*$
measured in the kaon rest frame.
$K^0_{e3\gamma}$ amplitude has infrared singularity  
at $E_{\gamma}^*\rightarrow0$, and a collinear singularity at
$\theta_{e\gamma}^*\rightarrow0$ when $m_e = 0$.
For
this measurement and the corresponding theoretical evaluation, we exclude
these regions by the restriction 
$E_{\gamma}^*>30$ MeV and $\theta_{e\gamma}^*>20^o$.


Two different theoretical approaches for evaluation of the branching ratio 
have been used. Current algebra technique together with the 
Low theorem were applied 
by Fearing, Fischbach and Smith (called FFS hereafter) \cite{FFS},\cite{FFS_new} and by 
Doncel \cite{Doncel}.
ChPT calculations were performed in \cite{Holstein}, \cite{ChPT} and
are being continuously improved \cite{ChPT_new}, \cite{Gasser}. 
The  ratio of the $K^0_{e3\gamma}$ to $K^0_{e3}$ decay probabilities, 
applying the
standard cuts on $E_{\gamma}^*$ and $\theta_{e\gamma}^*$, is
predicted to be between $0.95$ and $0.99\%$.
The amounts of direct emission in these various calculations differ, and are
roughly estimated
to be between 0.1 
and 1 \% of the size of the IB component. 

Two experimental measurements of the  $K^0_{e3\gamma}$ branching ratio
have been published.
The NA31 experiment obtained $Br(K^0_{e3\gamma},E_{\gamma}^*>30~ \rm{MeV}, 
\theta_{e\gamma}^*>20^o)/Br(K^0_{e3})=(0.934\pm0.036^{+0.055}_{-0.039}) \%$
\cite{NA31}.
The KTeV experiment gave a compatible value of the ratio 
$Br(K^0_{e3\gamma},E_{\gamma}^*>30~ \rm{MeV},\theta_{e\gamma}^*>20^o)/Br(K^0_{e3})=(0.908\pm0.008^{+0.013}_{-0.012}) \%$ \cite{KTeV}.
However, this value does not agree well with
theoretical predictions.


\section{Experimental setup}
The NA48 detector was designed for a measurement of direct CP
 violation in the $K^0$ system. Here we use data from a
dedicated run in September 1999
where a  $K_L$ beam was produced by 450 $\rm{GeV}/c$ protons from the
 CERN SPS incident on a beryllium target.
The decay region is located
120 m from the $K_L$ target after three
collimators and sweeping magnets. It is
 contained in an evacuated tube, 90 m long, terminated by a thin
 ($3\cdot 10^{-3} X_0$) kevlar window.

The detector components relevant for this measurement include
 the following:

The  { \bf magnetic spectrometer} is designed to measure the momentum of
charged particles with high precision. The momentum resolution  is  given by
\begin{equation}
\frac{\sigma(p)}{p} = \left( 0.48 \oplus 0.009 \cdot p \right) \%
\end{equation}
where $p$ is in $\rm{GeV}/c$. The  spectrometer consists of four drift chambers (DCH), each with 8 planes of sense wires
oriented along the projections $x$,$u$,$y$,$v$, each
one rotated by 45 degrees with respect to the previous one.
The spatial resolution achieved per projection is
$\rm{100~ \mu m}$ and the time resolution is $\rm{0.7~ns}$.
The volume between
the chambers is
filled with helium, near atmospheric pressure. The spectrometer
magnet is a dipole
with a field integral of 0.85 Tm and is placed after the first two chambers.
The distance between the first and last chamber is 21.8~m.

 The {\bf hodoscope} is placed downstream of the last drift chamber. It
consists of two planes of scintillators segmented in horizontal and vertical
strips and arranged in four quadrants. The signals are used
for a fast coincidence of two charged particles in the trigger. The time
resolution from the hodoscope is $200~\rm{ps}$ per track.

The  {\bf electromagnetic calorimeter} (LKr) is a quasi-homogeneous calorimeter based on liquid krypton,
with tower read out. The 13212 read-out cells have cross sections of
2 x 2 cm$^2$.
The electrodes extend from the front to the back
of the detector in a small angle accordion geometry.
The LKr calorimeter measures the energies of the $e^{\pm}$ and $\gamma$ quanta by 
gathering the ionization from their electromagnetic showers.
The energy resolution is :
\begin{equation}
\frac{\sigma(E)}{E} = 
\left(\frac{3.2}{\sqrt{E}}\oplus\frac{9.0}{E}\oplus0.42\right) \%
\end{equation}
where $E$ is in GeV, and the time resolution for showers with
energy between 3 GeV and 100 GeV is $500~\rm{ps}$.


The {\bf muon veto system} (MUV) consists of three planes of
scintillator counters, shielded by iron walls  of 80~cm thickness.
It is  used to reduce the $K_L \rightarrow \pi^{\pm}\mu^{\mp}\nu$
background.


Charged decays were triggered with a two-level trigger system. The trigger 
requirements were two charged particles in the scintillator hodoscope 
or in the drift chambers coming from the vertex in the decay region.

A more detailed description of the NA48 setup can be found elsewhere
\cite{NA48}.

\section{Analysis}

\subsection{Event selection }
The data sample consisted of about 2 TB of data from 100
million triggers, with approximately equal amounts recorded
with alternating spectrometer magnet polarities.
These data are the same which were used for the measurement of the 
$K_{e3}$ branching ratio \cite{BR}.
The following selection criteria were applied to the
reconstructed data
to identify $K_{e3}$ decays and to reject background,
keeping in mind the
main backgrounds to $K_{e3}$, which are
$K_L \rightarrow \pi^{\pm}\mu^{\mp}\nu$ ($K_{\mu3}$) and
$K_L \rightarrow \pi^+\pi^-\pi^0$ ($K_{3\pi}$):

- Each event was required to contain exactly two tracks,
of opposite charge,
and a reconstructed vertex in the decay region.
To form a vertex, the
closest distance of approach
between these tracks had to be less than 3~cm.
The decay region was defined by
requirements that the vertex had to be between 6 and 34~m from
the end of the last collimator
and that the transverse distance between the vertex and the
beam axis
had to be less than 2~cm.
These cuts were passed by 35 million events.

- The time difference between the tracks was required to be
less than $6~\rm{ns}$.
To reject muons,
only events with
both tracks inside the detector acceptance and without
in-time hits in
the MUV system were used.
For the same reason only
particles with a momentum
larger than 10~GeV 
were accepted.
In order to
allow a clear separation of pion and
electron showers,
we required the distance between the entry points of the
two tracks at the front face of the LKr Calorimeter 
to be larger than 25~cm. As a result 14 million events remained.

\begin{figure}[h]
\begin{center}
\epsfig{file=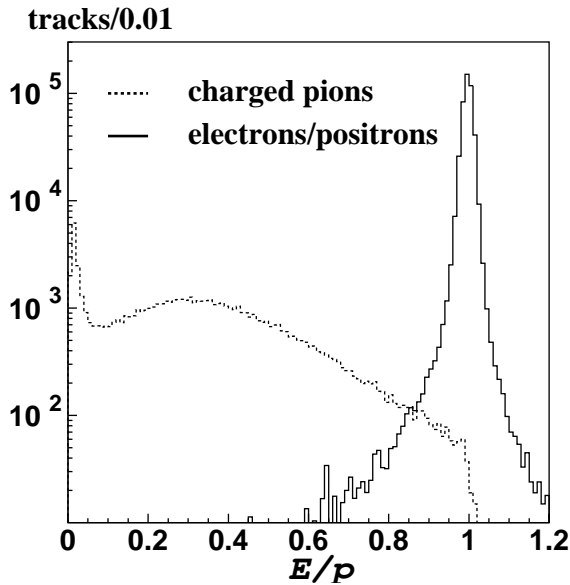,width=8.0cm,height=8cm}
\end{center}
\caption{Distribution of the ratio of the shower energy $E$ 
reconstructed by the LKr and the momentum $p$ reconstructed by
the spectrometer, for pions (dotted) and electrons (line) 
from $K_{e3}$ events (see text).}\label{eovp}
\end{figure}

- For the identification of electrons and pions, we used the ratio of the
 measured cluster energy, $E$, in the LKr calorimeter associated to a
track to the momentum, $p$, of this track as measured in the magnetic
spectrometer. The ratio $E/p$  for 
a sample of  75 000 pion tracks,
selected by requiring the other track of a 2-track event to be an 
electron with  $E/p > 1.02$,
 is shown in fig. \ref{eovp}. As a cross-check pion samples from
$K_{2\pi}$ and $K_{3\pi}$ decays were 
selected giving similar results.
Also shown in the figure is the distribution
for 450 000 electron tracks which are selected 
from 2-track events where  the
other track is a pion, with  $0.4<E/p<0.6$.  
For the selection of $K_{e3}$ events, we require one track to have
$0.93 < E/p < 1.10$ (electron) and the other track to have
$E/p < 0.90$ (pion). 11.7 million events were accepted.

- In order to reduce background from $K_{3\pi}$ decays,
we required the
quantity
\begin{equation}
 {P_0^{\prime}}^2=\frac{(m_K^2 -m_{+-}^2 -m_{\pi^0}^2)^2-
  4(m_{+-}^2m_{\pi^0}^2+m_K^2p_{\bot}^2)}{4(p_{\bot}^2+m_{+-}^2)}
\end{equation}
to be less than $-0.004 (\rm{GeV}/c)^2$.
In the equation above, $p_{\bot}$ is the transverse momentum of the two
 track system  (assumed to consist of two charged pions)
relative to the  $K_L^0$ flight
direction and   $m_{+-}$ is the
invariant mass of the charged system.
The variable ${P_0^{\prime}}^2$ is positively defined 
if the charged particles are pions from the decay $K_{3\pi}$ and its 
distribution has maximum at zero.
The cut removes $(98.94\pm0.03)\%$
of  $K_{3\pi}$
 decays and $(1.03\pm0.02)\%$ of
$K_{e3}$ decays as estimated with the Monte Carlo simulation (sect. 3.3).
After this cut, we were left with
11.4 million $K_{e3}$ candidate events.

The neutrino momentum in $K_{e3}$ decays
is not known and
the kinematic reconstruction of the kaon momentum from the measured
track momenta leads to a two-fold ambiguity
in the
reconstructed kaon momentum.
The solution with larger energy we call ``first
solution".
%
In order to measure the kaon momentum spectrum, we selected
events in which both solutions for the kaon momentum
 lie in the same bin of width 8 GeV. These $4\cdot10^5$ events we call
"diagonal events".

The last selection criterion was the
requirement that each of the two solutions for the kaon energy
had to be in the energy range (60,180) GeV.
 As a result of this selection,
$5.6 \cdot 10^6$
fully reconstructed  $K_{e3}$ events were selected from the total sample.
These selected events include radiative $K_{e3}$ events.

\begin{figure}[h!]
\begin{center}
\epsfig{file=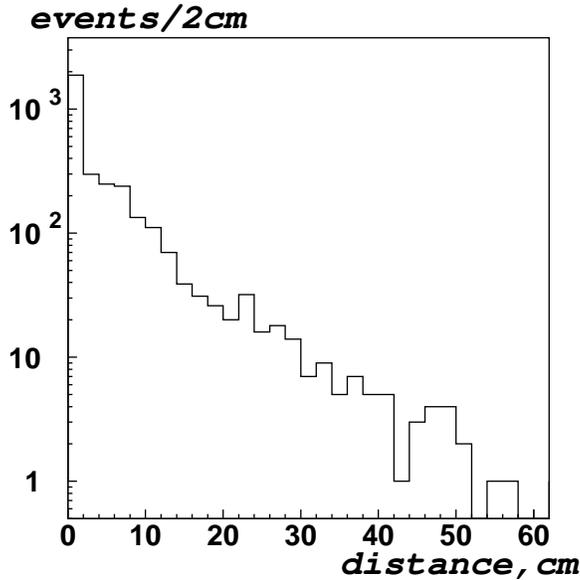,width=8.0cm,height=8cm}
\end{center}
\caption{Transverse distance between the pion entry point in
LKr and the position of a cluster
induced by pion interactions with matter; the pions here are selected from
$K^0_L\rightarrow\pi^{+}\pi^{-}$ decays where the 
entry points of the two tracks in the
LKr calorimeter are at least 80 cm from each other; clusters have a minimum 
energy of 4 GeV }\label{piclusters}
\end{figure}

For the selection of $K_{e3\gamma}$ events, 
the following additional requirements were made:

 The distance between the $\gamma$  cluster and the pion track in 
LKr  had to  be larger than  55 cm in order to 
allow a clear separation of the $\gamma$ cluster from 
pion clusters. 
As is shown in fig.  \ref{piclusters} the hadron showers 
 can extend over lateral distances of up to 60 cm 
from the track entry point in LKr.
After the requirements for $E_{\gamma}^*>30~\rm{MeV}$ and 
$\theta_{e\gamma}^*>20^o$ (for both solutions of the kaon energy), 
22 100  events survived.
To distinguish the $\gamma$ from the electron cluster we required the 
transverse distance
between the $\gamma$  cluster candidate and the electron track in
LKr to be greater than 6 cm. 
The electromagnetic 
transverse rms shower width in LKr is 2.2~cm.
An event was rejected if  
the $\gamma$ cluster candidate was less than 
16 cm away from the beam axis, because of the beam 
hole in the LKr calorimeter.
We also rejected events 
with a $\gamma$ cluster candidate with energy below 4~GeV because the 
energy resolution deteriorates below this threshold.
%
Finally an event was 
rejected  if the $\gamma$  was not in-time (more than $6~\rm{ns}$ 
time difference) 
with the associated cluster(s).
These cuts provided a sample of 
19 117 $K_{e3\gamma}$ candidates.

\subsection{Backgrounds}


The amount of background  was evaluated 
using a Monte Carlo simulation for other kaon  decays. 

The background to $K_{e3\gamma}$ events 
is small and comes from three sources -
$K_{3\pi}$ and 
$K_L \rightarrow \pi^{0}\pi^{\pm}e^{\mp}\nu$ ($K_{e4}$) 
decays as well as
$K_{e3}$ decays with an accidental photon. The $K_{3\pi}$ background was 
reduced 
by the cut on the variable
${P^{\prime}_0}^2$ and the electron identification through the $E/p>0.93$ 
condition. 
Variations of these cuts have a negligible effect, since the
probability to misidentify a pion for
an electron is only 0.57\% from fig. \ref{eovp}, and 
the  ${P^{\prime}_0}^2$ distribution is well reproduced by
the MC simulation. 
The estimated number of background 
events was $40^{+60}_{-40}$ events.
 
The $K_{e4}$ 
background was evaluated to be  $80\pm 40$ events
from the measured 
branching ratio and the calculated acceptance for these decays. 

The contamination from $K_{e3}$ decays with an accidental photon 
was estimated using 
the distribution of the time difference between the $\gamma$  cluster candidate
and the (average) time of the other cluster(s).
The number of events in the two control regions 
$(-25,-10)~\rm{ns}$ and $(10,25)~\rm{ns}$ were extrapolated to the signal  
region $(-6,6)~\rm{ns}$. 
The final number for this source
of background was estimated to be $20^{+40}_{-20}$ events, 
assuming a flat distribution. 


All backgrounds to $K_{e3\gamma}$
 add up to $140\pm82$ events or $0.7\%$ of the total $K_{e3\gamma}$
 sample of 19117 events.

  The main background to the normalization channel $K_{e3}$ arises from  
$K_{3\pi}$ and 
$K_{\mu3}$ decays.
The estimations were made as in
the case of  $K_{e3\gamma}$. 
All the  background decays together 
gave a $K_{e3}$ signature in less than  $9 \cdot 10^{-5}$ of the
cases ($<500$ events). This percentage is negligible compared to 
background sources in $K_{e3\gamma}$ decay.



\subsection{Monte Carlo Simulation}

\begin{figure}[t]
\begin{center}
\epsfig{file=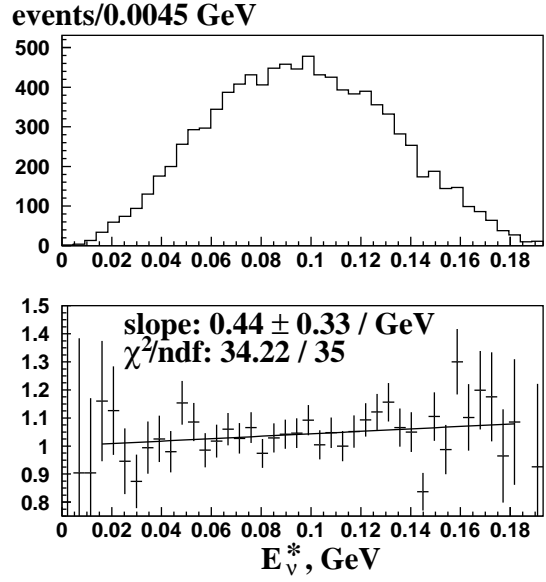,width=8.0cm,height=8cm}
\end{center}
\caption{Reconstructed neutrino energy in the 
Center of Mass System; upper part - 
experimental data distribution, lower part -  
normalized to unity ratio of
DATA/MC linearly fitted}\label{e_nu}
\end{figure}

\begin{figure}[h]
\begin{center}
\epsfig{file= 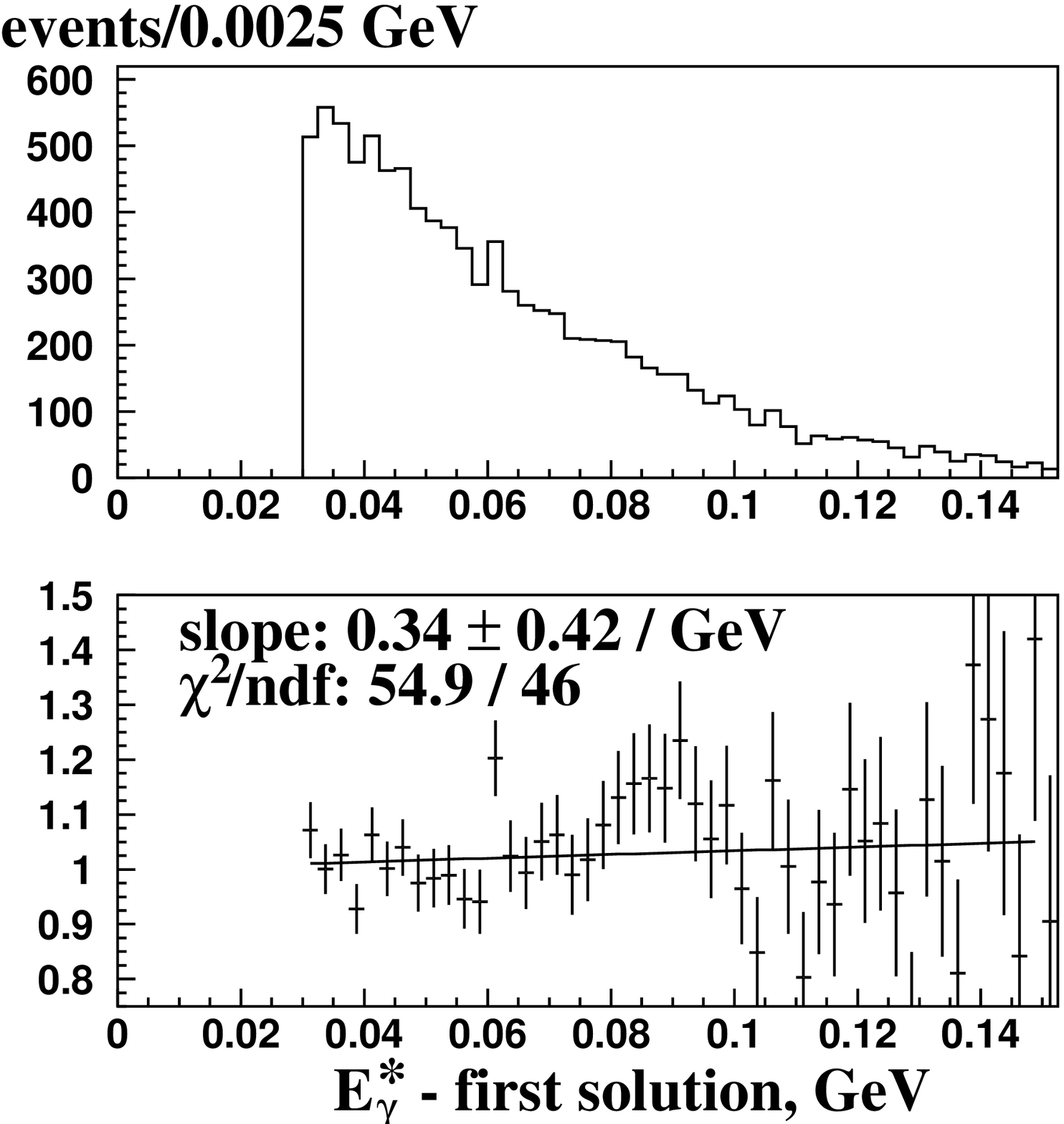,width=8.0cm,height=8cm}
\end{center}
\caption{First solution for $E_{\gamma}^*$; upper part - 
experimental data distribution, lower part - normalized to unity  ratio of
DATA/MC linearly fitted}\label{e1}
\end{figure}


\begin{figure}[h]
\begin{center}
\epsfig{file= 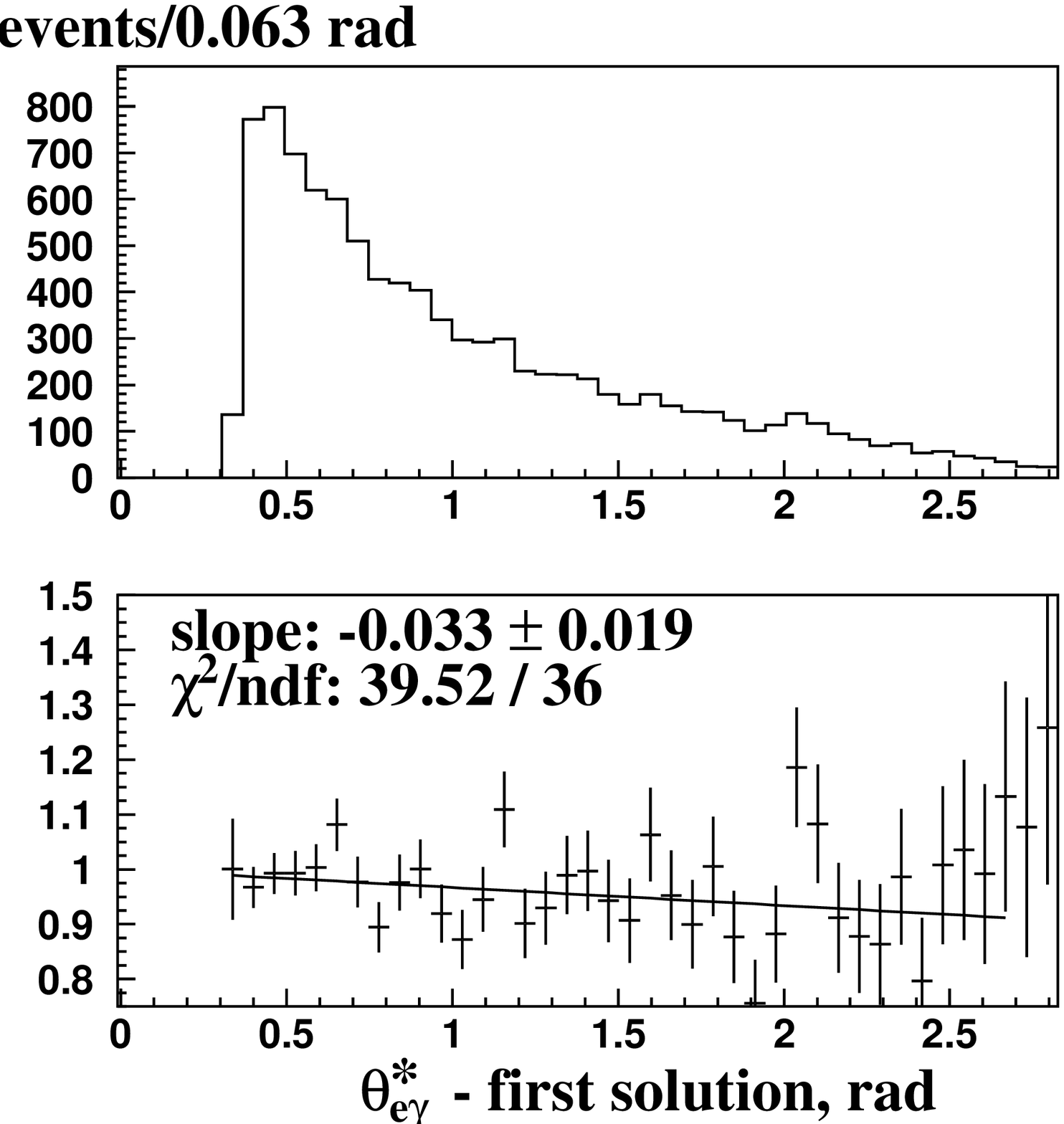,width=8.0cm,height=8cm}
\end{center}
\caption{First solution for $\theta_{e\gamma}^*$; upper part - 
experimental data distribution, lower part -  normalized to unity ratio of
DATA/MC linearly fitted}\label{cos1}
\end{figure}


In order to calculate the geometrical and kinematical acceptance of the 
NA48 detector,  a GEANT-based  simulation 
was employed \cite{NA48}. The kaon momentum spectrum from sect. 3.1 was 
implemented into the MC code. 
 The radiative corrections (virtual and real) were taken into account
by modifying the PHOTOS \cite{PHOTOS} program
package
in such a way as to reproduce the experimental data. This was achieved
by weighting the angular distribution $\theta^*_{e\gamma}$ in the 
centre-of-mass frame
such as to fit the experimental data (model independent analysis). 
With  this procedure the  MC and experimental data showed good agreement.
As an  example,
the distributions of
the neutrino 
energy, $\gamma$  energy 
and 
$ \theta_{e\gamma}^*$ (first solutions)
in the centre-of-mass frame are presented in 
figures \ref{e_nu}, \ref{e1},   
and \ref{cos1} 
respectively.
The upper 
plots of the figures show
 the experimental data
distributions and 
the lower show the ratio of the data and the MC spectra, normalized to 
unity. 
The plots represent data with the negative  
magnet polarity  and after the $K_{e3\gamma}$ selection.



The MC data were treated exactly in the
 same way as the  experimental data and were used for acceptance calculations.
The acceptance for $K_{e3\gamma}$ is 
$\epsilon(K_{e3\gamma})=(6.08\pm0.03)\%$
 as compared to the $K_{e3}$ acceptance 
$\epsilon(K_{e3})=(17.28\pm0.01)\%$. 



\subsection{Reconstruction and analysis technique}
We used the ``diagonal events''  
to measure the kaon momentum spectrum from $K_{e3}$ decays. 
However, as this reduces the data sample significantly, 
for the analysis of the branching ratio 
the problem was dealt with  in another way.
In the $K_{e3}$ selection it was required that both solutions were
 in the range (60,180) 
GeV. 
Further in the $K_{e3\gamma}$ selection 
events were rejected  if (at least) one of the two solutions for $E_\gamma^*$ was less
than $30$ MeV or  (at least) one of the two solutions for $\theta_{e\gamma}^*$
 was less than $20^o$.
The same procedure was used for selecting MC events when calculating
the acceptance.

An important issue are radiative corrections. Only the inclusive rate
($K_{e3\gamma}$ plus any number of radiative photons) is finite and calculable.
In our selection we have required only one hard $\gamma$ satisfying 
$E_{\gamma}^*>30~\rm{MeV}$ and 
$\theta_{e\gamma}^*>20^o$. 
In this way in the final selection events with one "hard" $\gamma$ and
any number soft photons are included. Events with two or more hard photons
are rejected. This loss has to be taken into account by MC in the
calculation of the corresponding acceptance. In order to check the MC we
have
compared the number of $\gamma$ clusters in the LKr calorimeter predicted
by the MC with the one in the experimental data. A slight difference
has been observed leading to a small correction of 0.05\% to the branching 
ratio. 
We
take this into account by a correction factor $C_M=0.9995$ to the
branching ratio.
Additionally we have
reanalysed our data, requiring at least one hard photon i.e.
accepting any number of photons. This is the inclusive rate which 
is finite and can be calculated. The result for $R$ agreed within 
0.2\% with the analysis requiring exactly one hard photon.





The trigger efficiency was measured to be 
$(98.1\pm0.1)\%$ for
$K_{e3}$ decays and $(98.1\pm0.6)\%$ for $K_{e3\gamma}$ decays.

On the basis of 19117  $K_{e3\gamma}$ candidates with an estimated
    background of 140 $\pm$ 82 events
 and 5.594 million $K_{e3}$ events (including
additional photons) after  background 
subtraction, and using the calculated acceptances,
the branching ratio was computed from
the relation: 

\begin{equation}
R=Br(K^0_{e3\gamma},E_{\gamma}^*>30~ \rm{MeV},\theta_{e\gamma}^*>20^o)/
Br(K^0_{e3})=
{N(K_{e3\gamma})Acc(K_{e3})\over N(K_{e3})Acc(K_{e3\gamma})} \cdot C_M
\end{equation}



The result from
9361 $K_{e3\gamma}$ events and 2.728 million $K_{e3}$ events for 
positive magnet
polarity was $R=(0.953\pm0.010)\%$ and from 9616 $K_{e3\gamma}$ 
events and 2.866
million $K_{e3}$ events for negative polarity, $R=(0.975\pm0.010)\%$, 
where the
errors are statistical. We now turn to the systematic uncertainties.

\subsection{Systematic uncertainties}

\begin{figure}[t!]
\begin{center}
\epsfig{file=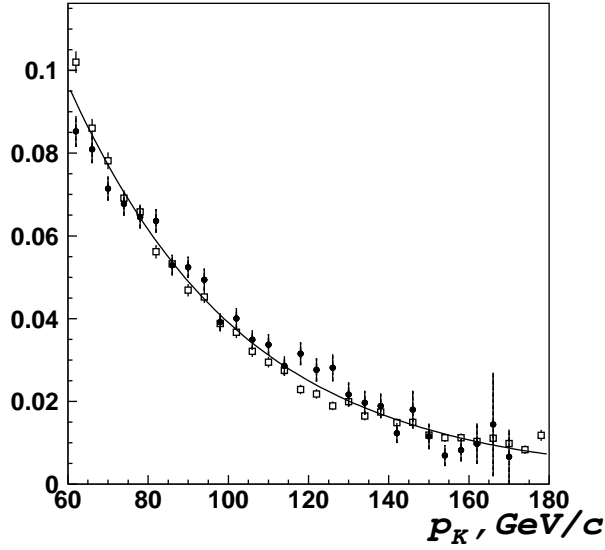,width=8.0cm,height=8cm}
\end{center}
\caption{Kaon momentum distribution obtained from $K_{e3}$ (line), 
$K_{2\pi}$ (open squares) and
 $K_{3\pi}$ (circles) decays. Arbitrary units on Y-axis}\label{specs}
\end{figure}

Our investigation of possible systematic errors showed that the
 biggest uncertainty comes from the kaon momentum spectrum. In order to
 determine the influence of this factor we reconstructed the
 experimental kaon momentum distribution from  
 $K \rightarrow \pi^+\pi^-$
and  $K \rightarrow \pi^+\pi^-\pi^0$ decays and implemented them 
in the MC simulation. 
The shape of the spectrum for the three decays is  shown in fig.
 \ref{specs}. 
The systematic error from the momentum
spectrum was estimated by taking the 3 different momentum spectra and
calculating the effect of this variation on the acceptance ratio of $K_{e3}$ 
and $K_{e3\gamma}$. It resulted in an relative uncertainty of 
$(~^{+6}_{-3}) \cdot 10^{-3}$.


The stability of the result upon the various cuts used
in the $K_{e3\gamma}$ selection was also investigated. 
The cuts were varied in between values 
which rejected no more than 10\% of the events. The biggest 
fluctuations in the 
branching ratio estimation 
were taken as systematic errors, and all the 
 errors were added in quadrature 
with a relative result of $\pm 5 \cdot 10^{-3}$.
   
Uncertainties  in accidental photon events and in other
background contributions are dominated by statistics and are not amongst
the largest of the systematic errors ($(~^{+2}_{-1}) \cdot 10^{-3}$ and
$(~^{+4}_{-3}) \cdot 10^{-3}$ correspondingly).
The influence of the $K_{e3}$ selection cuts 
to the final result  was estimated as in the case of $K_{e3\gamma}$ 
selection cuts. 
The quadratic addition of 
all these relative errors from variations of individual selection cuts 
yielded an inclusive relative error of $\pm 5 \cdot 10^{-3}$.
 The value of the form-factor $\lambda_+$ in the $K_{e3}$ decay was 
varied between 0.019 and 0.029.
The largest  fluctuation was taken as a relative systematic error - 
 $\pm 1 \cdot 10^{-3}$.

Our estimate of the systematic errors  is summarized in Table \ref{syst}.

\section{Results and conclusion}
\begin{table}[t]
\begin{center}
\mbox{
\begin{tabular}{||l|l||}
\hline
Source     &$\triangle R/R$    \\
\hline
$K_L$ spectrum&$~^{+6}_{-3}\cdot 10^{-3}$ \\
$K_{e3\gamma}$ selection  & $\pm 5\cdot 10^{-3}$   \\
$\gamma$  accidentals&$~^{+2}_{-1}\cdot 10^{-3}$ \\
Background uncertainties&$~^{+4}_{-3}\cdot 10^{-3}$   \\
$K_{e3}$ selection &$\pm 5\cdot 10^{-3}$   \\
Form-factor uncertainties &$\pm 1\cdot 10^{-3}$  \\
\hline
TOTAL & $~^{+11}_{-~9}\cdot 10^{-3}$\\
\hline
\end{tabular}
}
\end{center}
\caption{Relative systematic uncertainties to the Branching ratio}\label{syst}
\end{table}

\begin{figure}[b!]
\begin{center}
\epsfig{file=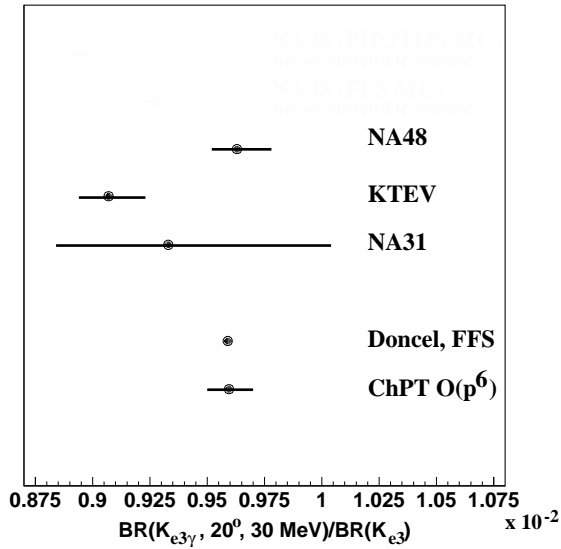,width=8.0cm,height=8cm}
\end{center}
\caption{Theoretical and experimental results for the radiative $K_{e3}$
branching ratio. The two lower entries in the plot are theoretical results.}\label{rescompar}
\end{figure}

The results are  based on 
$18977$ $K_{e3\gamma}$ and $5.594  \cdot 10^6$ $K_{e3}$ events.
We obtain the following
value for the branching ratio including the systematic error:
\begin{equation}
R=(0.964\pm0.008^{+0.011}_{-0.009})\%= (0.964^{+0.014}_{-0.012})\%
\end{equation}

%
%
Figure \ref{rescompar} shows this branching ratio compared to     
theoretical
and experimental results. The authors of ref. \cite{Gasser} have undertaken a 
serious effort to estimate the theoretical uncertainties in $R$,
while for the earlier theoretical values, this error is not known. 
These authors obtain $R = (0.96\pm0.01)\%$.
It appears
that our experimental result agrees well with the theoretical calculations
\cite{FFS_new}, \cite{Doncel}, including the most recent one \cite{Gasser}.
However our result is at variance with a recent experiment with similar 
statistical sensitivity \cite{KTeV}. Our measurement, with
a 1.5\% precision, therefore confirms the validity of calculations based
on chiral perturbation theory.

%


\section{Acknowledgement}

We would like to thank  Drs. J\"uerg Gasser, 
Bastian Kubis and Nello Paver for fruitful
discussions and for communicating to us their result in ref.[7] prior to
publication. We also thank the technical staff of the participating
institutes and computing centres for their continuing support.

\end{document}